\documentclass{osa-article}

\journal{osajournal}

\articletype{Research Article}

\usepackage{siunitx}

\begin{document}

\title{Total internal reflection based super-resolution imaging for sub-IR frequencies}

\author{Lauren E Barr,\authormark{1,*,+} Peter Karlsen,\authormark{1,+} Samuel M Hornett,\authormark{1} Ian R Hooper,\authormark{1} Michal Mrnka,\authormark{1} Christopher R Lawrence,\authormark{2} David B Phillips,\authormark{1} and Euan Hendry\authormark{1}}

\address{\authormark{1}Department of Physics and Astronomy, University of Exeter, Exeter, EX4 4QL, UK\\
\authormark{2}QinetiQ, Cody Technology Park, Ively Road, Farnborough, GU14 0LX, UK\\
\authormark{+} These authors contributed equally to the work\\}

\email{\authormark{*}l.barr@exeter.ac.uk} 



\begin{abstract}
For measurements designed to accurately determine layer thickness, there is a natural trade-off between sensitivity to optical thickness and lateral resolution due to the angular ray distribution required for a focused beam. We demonstrate a near-field imaging approach that enables both sub-wavelength lateral resolution and optical thickness sensitivity. We illuminate a sample in a total internal reflection geometry, with a photo-activated spatial modulator in the near-field, which allows optical thickness images to be computationally reconstructed in a few seconds. We demonstrate our approach at 140\,GHz (wavelength 2.15\,mm), where images are normally severely limited in spatial resolution, and demonstrate mapping of optical thickness variation in inhomogeneous biological tissues.
\end{abstract}

\section{Introduction}

Long wavelength measurements such as Terahertz (THz) time domain spectroscopy are often used as a route to film thickness determination of visibly opaque materials, either using time of flight approaches \cite{Zhong2005,Abraham2010} or through spectral analysis \cite{Johnson2001,Nagatsuma2013}. These techniques can achieve optical thickness measurements of exceptionally high accuracy, sometimes approaching micron scale \cite{Johnson2001,Abraham2010}. However, to achieve good sensitivity to optical thickness conventionally requires illumination with near-collimated beams (i.e.\ only weakly focused beams with a narrow angular spectrum). This requirement limits the lateral spatial resolution of these methods, as high thickness sensitivity is not compatible with the strong focusing usually necessary to achieve high lateral resolution. This trade-off is also inherent in total internal reflection imaging techniques that have been demonstrated to date \cite{Wojdyla2013,Grognot2015}. The compromise between optical thickness sensitivity and lateral spatial resolution means that time domain techniques have been most successfully employed to investigate homogeneous samples such as polymer films \cite{Chen2019a} or crystalline pallets \cite{Ibrahim2019}. 

Malleable and inhomogeneous materials, such as biological tissues, are particularly difficult to characterise by conventional means \cite{Bowman2015}. Mechanical measurements of thickness are unreliable when the sample can be compressed, and time consuming if a map of varying thickness is required. Fully characterising a sample made of opaque layers of different properties and thicknesses poses an even greater challenge. Often, cross-sections of such samples are taken and studied by eye. However this is destructive, and does not give the three-dimensional full picture of an inhomogeneous sample. Imaging in the THz and mm-wave bands offers a promising alternative for imaging the structure of inhomogeneous, and particularly biological materials, where water absorption leads to a natural contrast mechanism \cite{He2017}. In addition, THz radiation is non-ionising, and hence is of particular interest for medical imaging. Recent examples of biological and medical THz imaging include the study of cartilage from the knee~\cite{Stantchev2018}, observation of the healing process of scar tissue~\cite{Fan2017}, measuring the hydration of corneas~\cite{Taylor2017} and imaging water-dense cancers in breast tissue~\cite{Yu2012,Ashworth2009,Fitzgerald2014}. However, due to the low resolution associated with diffraction limited imaging in these bands, the inherent sample inhomogeneity on sub-wavelength scales is a major challenge when imaging biological tissues \cite{Bowman2015}. To overcome this limitation, THz images of sub-wavelength surface features have been captured using near-field scanning techniques. However, weak signals typically emanating from  scanning probe tips, which are sub-diffraction limited in scale, coupled with noisy detectors, results in lengthy measurement times \cite{Zhao2014,Eisele2014,Blanchard2011}. Moreover, the sensitivity of detectors in the THz and mm-wave spectral bands is poor, as the low energies carried by photons makes their control and detection extremely demanding. The most sensitive detectors require deep cryogenic cooling, which is problematic for the development cameras consisting of multi-sensor arrays, as the individual signals from thousands of detectors must be acquired per frame.

Emerging computational imaging techniques offer a potential way forward by redefining our understanding of how images can be recorded. Such systems do not form an image directly, but offload aspects of the image formation process to a computer. `Single-pixel' computational imaging systems circumvent the need for detector arrays, and boost the magnitude of near-field measurements by recording signals from many locations at once. These systems encode spatial information in the temporal dimension, enabling operation at wavelengths where multi-pixel sensors are challenging to fabricate, prohibitively expensive, or simply do not exist \cite{Hack2016}. Images are reconstructed from measurements using a single-element detector, which itself records only the total incident intensity, in conjunction with a spatial light modulator (SLM). The SLM spatially modulates (i.e.\ varies) the intensity of light, and so enables sequential illumination of the target object with a series of patterned light fields, each of which probes a different subset of the spatial information in the scene. Recently there have been several demonstrations of deeply sub-wavelength single-pixel transmission imaging at THz frequencies~\cite{Chan2007,Shrekenhamer2013,Saqueb2016,Hornett2016}.

In this work we demonstrate a new form of millimetre-wave computational imaging which enables the determination of overlayer optical thickness with significantly sub-wavelength precision, while also retaining sub-wavelength lateral resolution. We refer to our approach as a type of total internal reflection (TIR) imaging: in the absence of sample-induced loss we anticipate perfect reflection of the illuminating beam, while lossy samples reduce reflection in a spatially varying way, producing sample dependent image contrast. Using mm-wave radiation in a TIR geometry, rather than optical frequencies as previously studied~\cite{Axelrod2016}, means the evanescent fields can probe millimetric distances into samples - distances relevant, for example, to the determination of the thickness of structural tissue changes found in cancerous samples \cite{Bowman2016}, or the study of the layered structure of the skin \cite{Pickwell2004}. Our imaging approach is based upon a bespoke optically addressed mm-wave spatial modulator, integrated with a dove prism in a TIR geometry. This allows dynamic patterning of the evanescent mm-wave field probing the underside of potentially non-transmissive objects~\cite{Martin-Fernandez2013}, granting access to a range of new types of sample, including biological tissues. This approach can capture images with sub-wavelength resolution for any sub-IR frequency, with an upper frequency determined by the plasma frequency of the modulator ($\sim 1 \times 10^{12}$ Hz). We show that the near-field placement of the modulator decouples the thickness sensitivity from lateral resolution, subverting the trade-off which normally limits far field imaging approaches. As a result, and in contrast to previous single-pixel THz imaging systems, which typically produce an absorption image of a thin sample in transmission mode, our total internal reflection geometry gives access to a richer vein of information about the sample: the reflected signal is highly sensitive to both material properties and sample thickness. To the best of our knowledge, this is the first time a near-field imaging approach has been combined with total internal reflection, yielding the best of both worlds: the high lateral resolution of near-field imaging, with the good thickness sensitivity of a TIR geometry.

Critical to the performance of our imaging system is the mm-wave spatial modulator, and we describe in detail how this has been tuned for use at mm-wavelengths. This optimisation balances the resolution, frame-rate and contrast of the modulator. Due to the TIR geometry, the resulting modulator is highly efficient, circumventing the requirement for high intensity femtosecond-pulse optical pump beams~\cite{Liu2016,Stantchev2020}, vastly simplifying the experimental set-up in comparison with earlier work \cite{Stantchev2016,Nozokido1997,Okada2011}, removing unwanted heating of sensitive samples, and reducing optical noise. We show that our TIR based imaging system offers a new way forward to simultaneously image and measure the thickness of biological samples. To interpret our images throughout our study, we compare them with a transfer matrix based model describing expected levels of reflection as a function of frequency, incident angle and material parameters of the sample. We investigate the effect of the sample characteristics on image formation, and experimentally show that the measured signal encodes information about both the material and the film thickness. In our proof-of-principle experiments, we show that TIR mm-wave images reveal regions of fat and protein in porcine tissue samples, highlighting the system as a good candidate for imaging water-dense samples such as cancer tumours. We also demonstrate that it is possible to detect objects embedded within a thick optically-opaque material, by imaging a metal target on the far side of a fatty layer of tissue. While at present we have implemented this technique for mm-wave imaging, the principles described here could be broadly applied from GHz through to THz frequencies.

\section{Designing a mm-wave TIR imaging system}

\begin{figure}[ht]
    \centering
    \includegraphics[width=8cm]{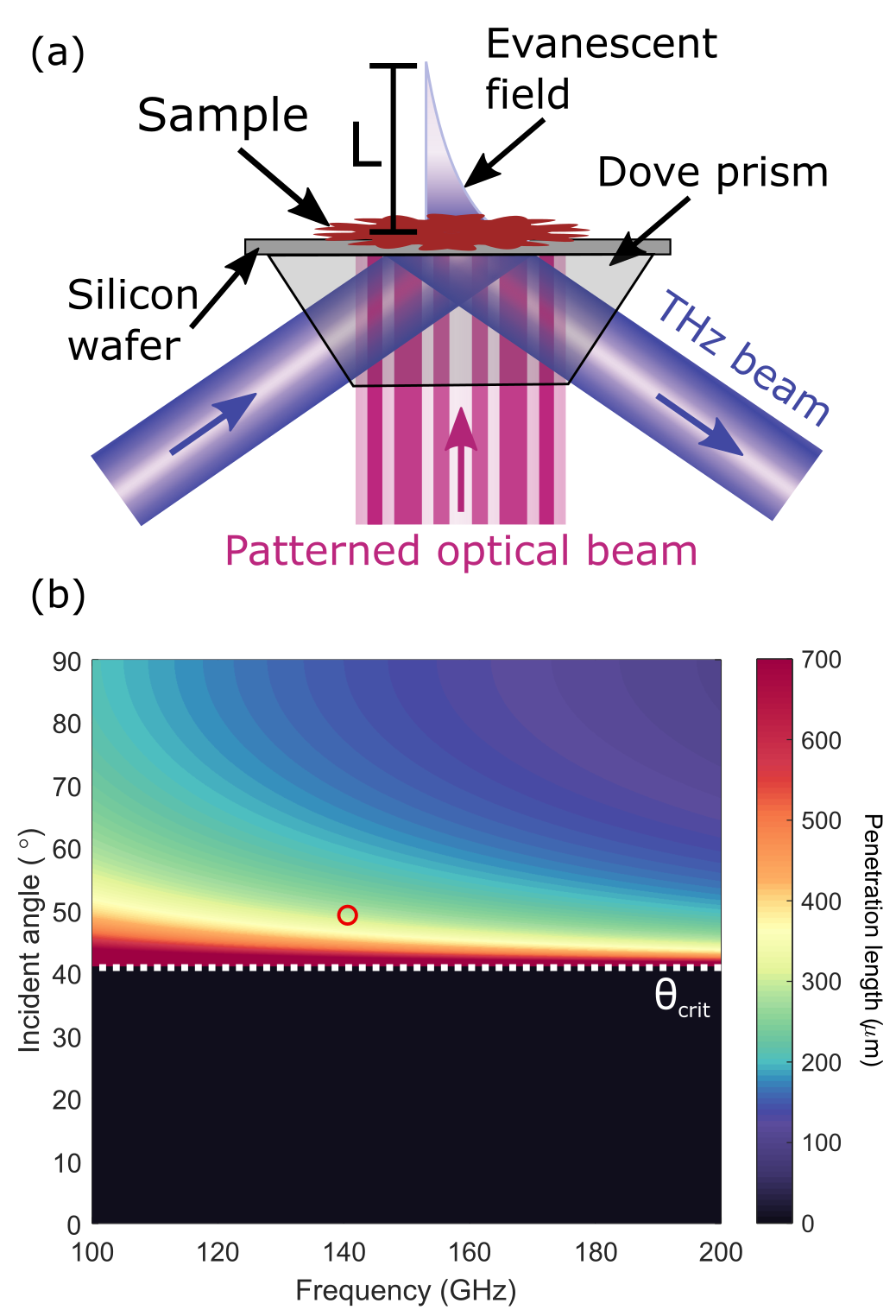}
    \caption{\textbf{The mm-wave imaging system.} (a) Schematic showing mm-wave modulator using total internal reflection. (b) Calculated penetration length of evanescent fields, L, at the silicon-air interface for a range of frequencies and angles of incidence. The \SI{390}{\micro\meter} thick silicon wafer has a permittivity of $11.7 + 0.003i$, and the prism medium has a permittivity of 2.49. The white dashed line shows the critical angle, below which there are no evanescent fields, and the red circle denotes the selected experimental parameters.}
    \label{fig_schem}
\end{figure}

Our TIR imaging system is based on all-optical modulation (known as photo-modulation), where a visible or near-IR light source is used to change the transparency of a photo-active medium to low frequency radiation. This approach is often chosen because it is relatively simple to implement, can be very fast and can operate over a broad range of mm-wave and THz frequencies \cite{Bai2015,Rahm2013}.

Figure \ref{fig_schem}(a) shows a schematic of our TIR imaging system. An IMPATT (IMPact ionization Avalanche Transit-Time) diode source ({\it TeraSense}, power 30 mW) produces a continuous wave with a frequency of 140\,GHz $\pm$ 1\,MHz (a wavelength of around 2.15\,mm). This passes through a dove prism made of a polymer, TPX (polymethylpentene) \cite{Inc2019}, which is transparent to both optical and GHz/THz radiation, with a refractive index of 1.58 and low loss (a tangent of 8$\times$10$^{-4}$) at 140\,GHz. The mm-wave spatial modulator is formed from a \SI{390}{\micro\meter} thick photo-active silicon wafer, placed on top of the prism, with a thin layer of immersion oil (refractive index of 1.513) between the wafer and prism to prevent air-gaps. The sample is then placed on top of the silicon wafer: its proximity to the mm-wave modulator means we can pattern the mm-wave on a sub-wavelength scale, as the fields at the surface of the silicon do not diffract. The specular mm-wave component that is totally internally reflected at the silicon-air interface is detected in the far field by a detector ({\it TeraSense}). 

Before photo-activation, the silicon wafer is largely transparent to mm-wave radiation. When mm-waves are incident at an angle greater than the critical angle, they are totally internally reflected from the silicon-air interface. Upon illumination of the silicon wafer with visible light, electron-hole pair photo-excitation increases the mm-wave absorption and so reduces mm-wave reflection. Therefore, by projecting optical binary intensity patterns at normal incidence onto the silicon wafer using light from a LED pump beam ({\it Thorlabs SOLIS}, wavelength 623\,$\pm$\,5\,nm, power 4.7\,W) patterned with a digital micro-mirror device (DMD, {\it Vialux}), we are able to spatially control which areas of the modulator are more reflective to mm-wave radiation. In this way, the pattern in the optical pump beam is dynamically imprinted onto the mm-wave field bathing the sample, enabling single-pixel imaging to be performed. The projected optical binary patterns are drawn from the Hadamard basis. This basis yields reconstructed images with significantly higher signal-to-noise ratio (SNR) compared to simple raster scanning of a transparent (or absorbing) window, and the SNR is uniformly distributed over the reconstructed image as the set is orthogonal~\cite{Sloane1976a,Decker1970}. Detail of the image reconstruction can be found in section 3.1 of the supplementary material. Our data acquisition rate is nominally 2 kHz, limited by the lifetime of the charge carriers in the silicon wafer. For $64 \times 64 = 4096$ pixels, we obtain images with sufficient signal to observe features in the samples in around 4 seconds. To allow for low noise image analysis, the images presented below have been averaged for several minutes (see section supplementary for a full analysis), which equates to 1 minute 22 seconds of exposure when omitting programming wait times. 

The penetration length of the evanescent field in air, L (eqn. (S1) in the supplementary material), depends on both the incident angle of the mm-wave beam and its frequency, as shown in Fig.~\ref{fig_schem}(b). L is maximised for low frequencies (i.e. longer wavelengths) and angles very close to the critical angle (white dotted line). These dependencies allow L to be tuned to match the sample characteristics. In this work we elected to image at a frequency of 140 GHz and incident angle of $49^{\circ}$ (denoted by the red circle in Fig.~\ref{fig_schem} (b)), giving a penetration depth in air of $\sim$\SI{300}{\micro\meter}.

\section{Responsivity to sample thickness}

\begin{figure}[htbp]
    \centering
    \includegraphics[width=8cm]{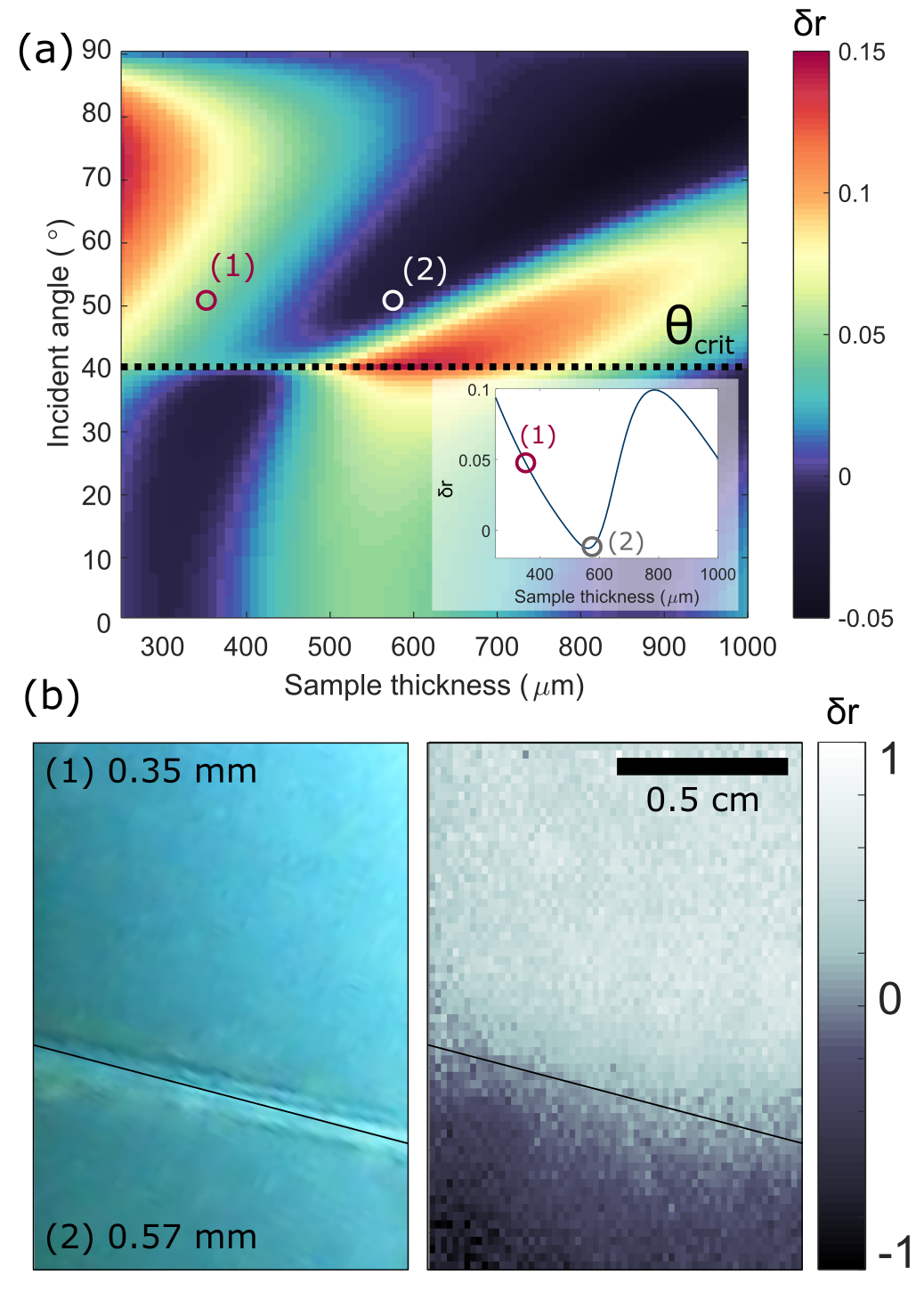}
    \caption{\textbf{Determining the thickness of a sample.} (a) Calculated $\delta r$ when a sample of permittivity $3.5 + 0.53i$ is present on top of the silicon wafer, as a function of sample thickness and incident angle of the mm-wave beam. The inset shows a slice through the plot at $50^{\circ}$ incidence. Results are calculated at 140 GHz frequency, and an incident photoexcitation wavelength of 623 nm and power of 220 Wm$^{-2}$ when illuminated, and 0 Wm$^{-2}$ when dark. The silicon wafer has a thickness of \SI{390}{\micro\meter} and a charge carrier lifetime of \SI{75}{\micro\second}. The critical angle is marked by the dotted black line. The red and white (inset: grey) circles (1) and (2) denote the parameters corresponding to the image in (b). (b) Optical (left) and mm-wave (right) images of a polymer sample with regions of two different thicknesses.}
    \label{fig_tape}
\end{figure}

When a layer of material is placed on top of the silicon wafer, the intensity of the mm-wave beam reflected back into the detector depends upon both the complex refractive index and thickness of the layer. If the refractive index of the material is less than that of the prism, one expects evanescent fields at the silicon-sample boundary that are sensitive to the properties of the sample \cite{Martin-Fernandez2013}. If the sample has an index higher than the prism, one breaks the total internal reflection condition at the silicon-sample interface, allowing a propagating wave to interrogate the sample layer.

To study these effects we image a step in a polymer sample of known height and permittivity (3.5 + 0.53i), and compare the image to the modulation predicted using the transfer matrix approach, as described in reference \cite{Hooper2019}. The sample under investigation is created by layering cellulose from tape to create films of different thicknesses. Reconstructed images are spatial maps of $\delta r(x,y)$, which is the difference in reflected signal at each spatial location ($x$, $y$), as the mm-wave modulator is switched from its Off (dark) to On (i.e.\ photo-activated) state. Figure~\ref{fig_tape}(a) shows the predicted level of mm-wave modulation $\delta r$ as a function of incident angle and sample thicknesses. Circles (1) and (2) indicate the expected signals at thicknesses of $350 \pm \SI{10}{\micro\meter}$ and $570\pm \SI{10}{\micro\meter}$. Figure \ref{fig_tape}(b) shows experimentally obtained visible (left panel) and mm-wave (right panel) images of the polymer step.

One intriguing feature of Fig.~\ref{fig_tape} is that the sign of $\delta r$ changes as a function of thickness. $\delta r>0$ indicates a decrease in reflection of mm-wave radiation on photo-excitation of the silicon wafer, while $\delta r<0$ indicates an increase. $\delta r<0$ is somewhat unexpected: why would an introduction of absorption into the silicon layer lead to an \textit{increase} in the level of reflected radiation? This surprising effect arises due to an optimal matching condition between the absorption and radiative loss channels of the system, which defines critical coupling conditions \cite{Haus1984}. This is similar to an effect observed for plasmonic cavities \cite{Herminghaus1994,Pockrand1977}, and arises here due to a Fabry-Perot resonance present in the sample itself. It is clear that the change in sign of $\delta r$ is both predicted in our wave modelling (Fig.~\ref{fig_tape}(a)) and observed in experiment (Fig.~\ref{fig_tape}(b)). However, the negative signals measured in experiment have a larger magnitude than those predicted from our modelling: we suspect this arises due to small air gaps in between cellulose layers changing the matching condition. Nevertheless, as demonstrated, an image contrast ($\delta r$) which \textit{varies in sign} upon a change in sample thickness of an order of magnitude smaller than the wavelength, has significant measurement potential, especially for challenging samples made from malleable and soft materials. Furthermore, we show, in section 4.5 of the supplementary material, that if we could take several images at different angles of incidence, it is possible to simultaneously extract both the thickness and complex refractive index of a sample using our technique. This is only possible due to the well defined incident angle in our measurement approach and is not usually possible in high resolution imaging, as the angular distribution required for a focused spot is incompatible with well defined angle dependencies. This issue is also often faced by e.g. time-of-flight measurements, where the delay of pulses depends on the optical path length, and it is not possible to separate this into contributions from the sample thickness and refractive index.

\section{Imaging biological tissue}

\begin{figure}[htbp]
    \centering
    \includegraphics[width=8cm]{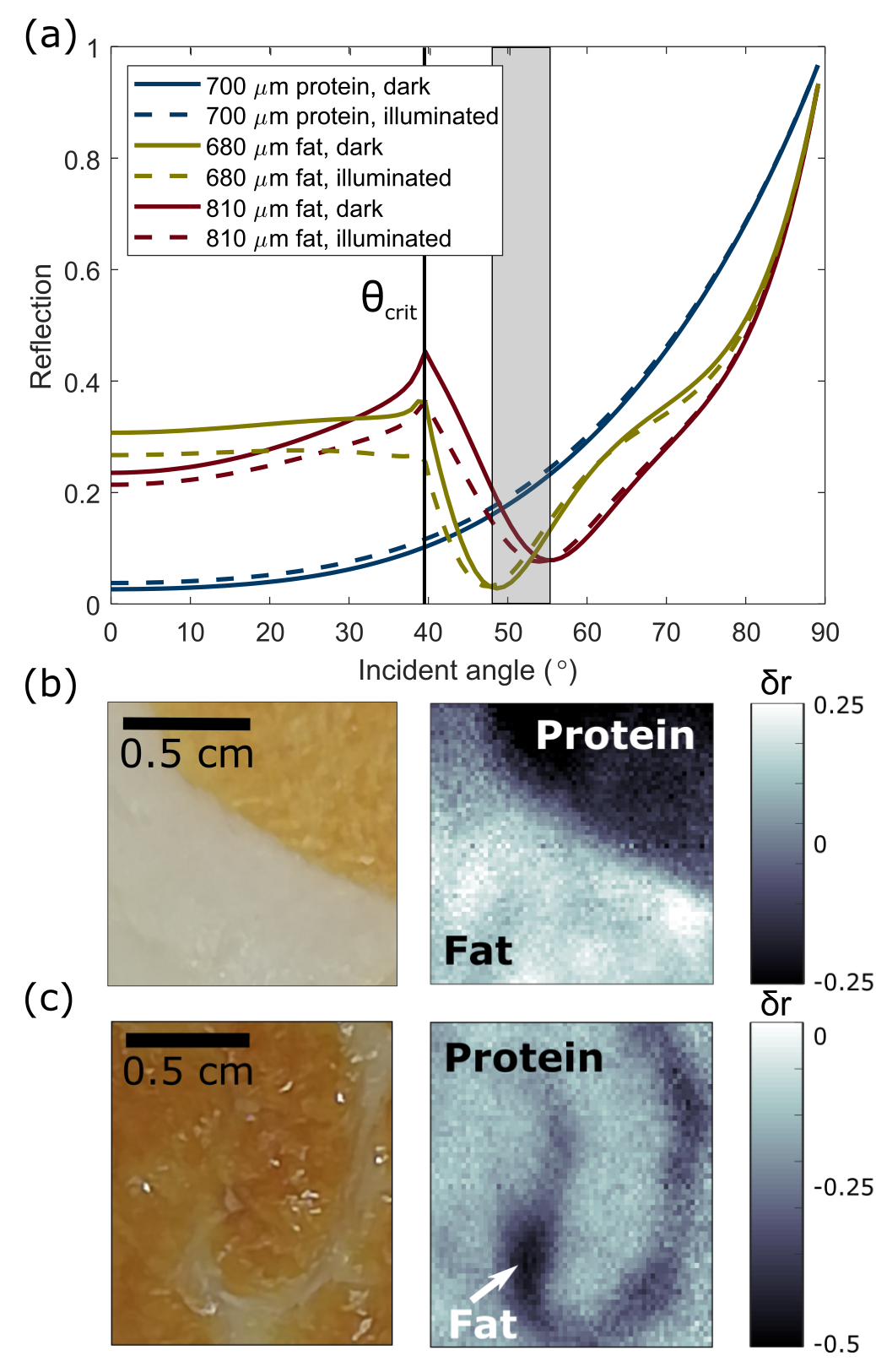}
    \caption{\textbf{Imaging biological tissue.} (a) Calculated reflection of a TE polarised mm-wave at 140 GHz when various biological tissues are placed on top of the silicon wafer, when the silicon is dark (solid lines) or illuminated (dashed lines) with a photoexcitation beam of 220 Wm$^{-2}$ at 623 nm wavelength. The silicon wafer has a thickness of \SI{390}{\micro\meter} and a charge carrier lifetime of \SI{75}{\micro\second}. The critical angle ($\theta_{\textrm{crit}}$, black vertical line) and the range of angles where the modulation is of opposite sign for the two different thicknesses of fat (grey shaded area) are marked on the plot. (b) and (c) images of porcine tissue samples at visible (left) and 140 GHz (right) frequencies. (b) shows a large fatty region well separated from the protein, and (c) shows thin filaments of fat embedded in protein, while the regions of fat and protein are clearly distinguishable in both. The 64 $\times$ 64 pixel mm-wave images are an average of 100 images taken with TE polarisation, where each took 4.1 seconds to collect (images collected over shorter times can be found in section 3.3 of the supplementary material).}
    \label{fig_bacon}
\end{figure}

We now demonstrate imaging of inhomogeneous biological tissues using our mm-wave TIR imaging system. The sample is a $750\pm \SI{250}{\micro\meter}$ thick piece of sliced porcine tissue, which exhibits strong spatial variation in properties, varying from fatty to protein rich regions. While one can expect a high degree of variation in the complex permittivities of biological tissues (see section 4.4 of the supplementary material) \cite{Ashworth2009,Bowman2016,Sun2009,Sun2011}, for now in our model we assume protein rich and fatty are described by $\epsilon_{\textrm{fat}} = 2.89 + 0.64i$ and $\epsilon_{\textrm{protein}} = 8.63 + 11.20i$, taken from reference \cite{Gabriel1996}. Using these values, Fig.~\ref{fig_bacon}(a) shows modelling of reflected mm-wave signal when a layer of fat or protein is placed on top of the wafer, as a function of incident angle, for both photo-activated (220 Wm$^{-2}$) and dark silicon. We show modelling results for three different thicknesses of material within the experimentally measured range.

Our model indicates that materials with high losses, such as protein, do not exhibit strong variation in $\delta r$ as sample thickness changes, due to the strong absorption of the mm-wave beam. We show the dependency of reflection on incident angle for a single thickness of protein tissue (\SI{700}{\micro\meter}, blue plot) in Fig.~\ref{fig_bacon}(a). We also see that the difference in predictions for an illuminated and non-illuminated modulator is very small (i.e. $\delta r$ is expected to be small) for protein tissue. There are two factors which determine this: firstly the absorption of the mm-wave is higher in protein than in fat, due to its higher water content; secondly, the real part of the refractive index of protein is close to that of silicon, and therefore a significant drop in the reflection is expected in this case. Fatty tissue, on the other hand, has a much lower water content, and hence absorption is lower. We predict fatty tissue will exhibit a strong variation of $\delta r$ on sample thickness. This is shown in the red and yellow plots in Fig.~\ref{fig_bacon}(a), which show reflection as a function of incident angle for two thicknesses (\SI{680}{\micro\meter}, yellow plot, and \SI{810}{\micro\meter}, red plot). Here we see that $\delta r$, the difference between dashed and solid lines, is expected to change sign when passing through the minimum in reflection at around $50^{\circ}$ ($60^{\circ}$) for a thickness of \SI{680}{\micro\meter} (\SI{810}{\micro\meter}). We therefore see a range of angles, marked by the grey region, were we expect the signal to be completely opposite in sign for a thickness difference of just \SI{130}{\micro\meter} (i.e. just 5$\%$ of the mm-wave wavelength).

In Figs.~\ref{fig_bacon}(b) and (c) we present mm-wave images of two different samples of porcine tissue: one with a large region of fat well separated from the protein rich region (b), and one with filaments of fat  on the scale of the imaging wavelength running through (c). In both images, the regions of fat and protein are clearly distinguishable, as seen by comparison to the optical images on the left. As predicted, $\delta r$ is close to zero within regions of protein in both Figs.~\ref{fig_bacon}(b) and (c). Approximate measurements using a set of Vernier Calipers gave the thickness of fatty regions in Fig.~\ref{fig_bacon}(b) to be $810\pm \SI{200}{\micro\meter}$, while the thickness of the fat in Fig.~\ref{fig_bacon}(c) was estimated to be $680\pm \SI{150}{\micro\meter}$. As predicted by our model, the measured $\delta r$ of fatty tissue regions in each image have opposite sign. We present further simulations in section 4.3 of the supplementary material that illustrate this sensitivity to permittivity and sample thickness.

The low absorption of fat also allows us to image objects behind layers of fatty tissue. This is shown experimentally in section 4.6 of the supplementary material, and demonstrates imaging through visibly opaque fatty tissues to yield information on the structure of a sample that would not otherwise be known.

\section{Exploring physical limitations of the system}

\begin{figure}[htbp]
    \centering
    \includegraphics[width=8cm]{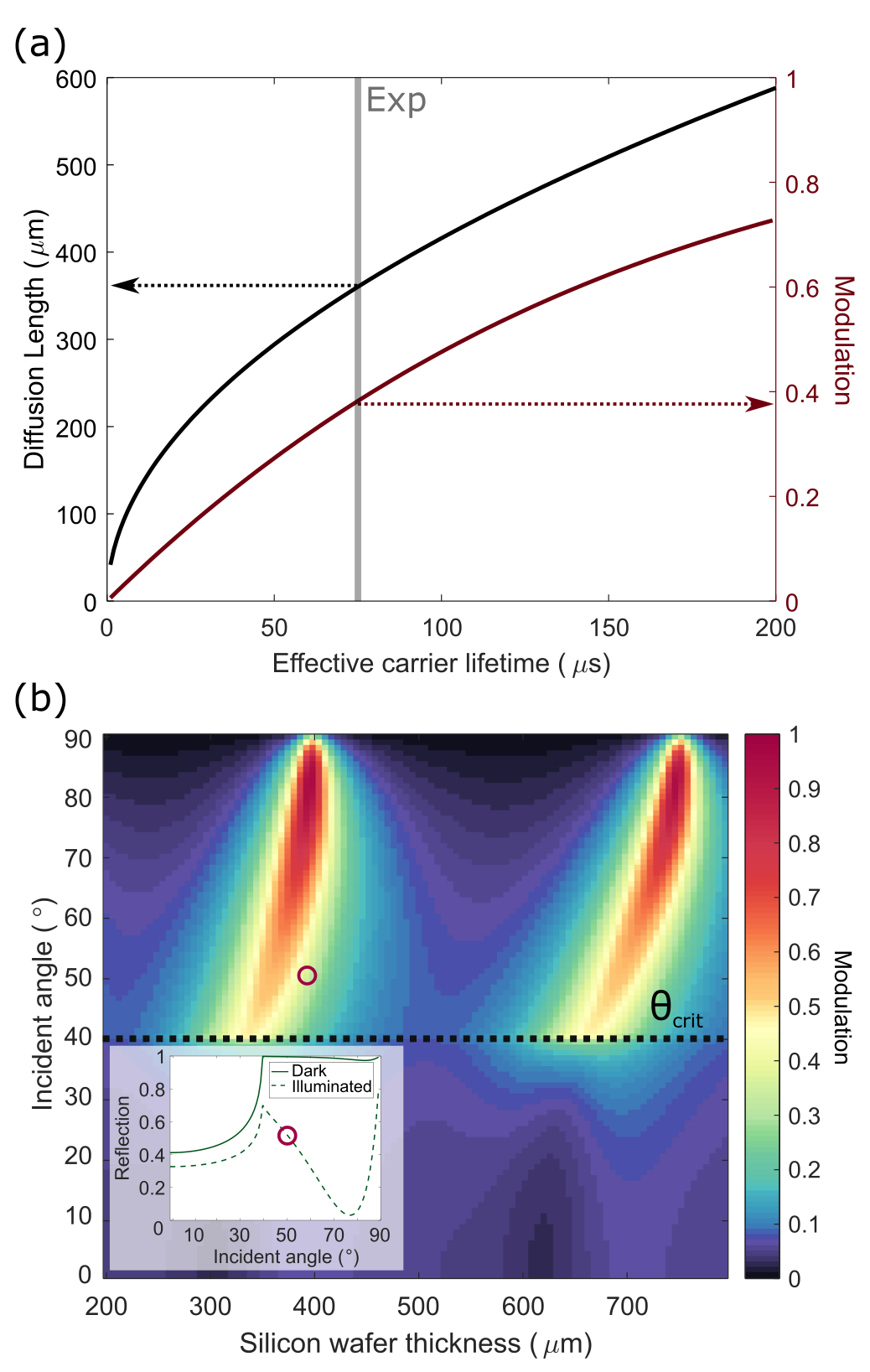}
    \caption{\textbf{Investigating design parameters.} (a) Calculated diffusion length of carriers (left, black axis) and modulation (right, red axis) as a function of effective charge carrier lifetime for a mm-wave beam incident at $50^{\circ}$ through the prism material onto a \SI{390}{\micro\meter} thick wafer. The grey line shows the lifetime of the wafer used in experiment. (b) TE polarised modulation of a mm-wave signal as function of angle of incidence in the prism material and silicon wafer thickness, assuming a charge carrier lifetime of \SI{75}{\micro\second}. Inset shows the reflection for dark and illuminated states for a \SI{390}{\micro\meter} thick wafer. All results are calculated at 140 GHz mm-wave frequency, and an incident photoexcitation wavelength of 623 nm and power of 220 Wm$^{-2}$ when illuminated, and 0 Wm$^{-2}$ when dark. The red circles show the experimental parameters.}
    \label{fig_mod}
\end{figure}

There are several inter-related factors that determine the performance of our TIR computational imaging system. Conventional diffraction limited imaging systems can reach resolutions of $\sim\lambda$/2NA, where NA is the numerical aperture defined by the distribution of incident rays. To achieve high depth sensitivity one requires illumination with near-collimated mm-waves: our system has a mm-wave NA = 0.06, similar to NAs used in both time of flight systems \cite{Abraham2010,Zhong2005} and previous TIR imaging systems \cite{Wojdyla2013,Grognot2015}. An NA = 0.06 corresponds to an expected diffraction limited resolution $\sim$1.8\,cm for 140 GHz - this demonstrates the trade off between lateral and depth resolution normally inherent to both time of flight and TIR approaches. However, in our design, the in-plane lateral resolution of images is determined by the local modulation of the incident mm-wave by our photomodulator. As shown in the supplementary information section 4.1, due to the low NA of our detector, our modulation signals are dominated by the specular reflection from the sample, which maintains the low angular dispersion of the measurement. Yet the specular reflection is still affected by local photomodulation, as can be seen by the small spatial features evident in some of our images. In this case, the image resolution will be determined by the optical pump beam, and subsequent charge carrier diffusion within the silicon wafer (i.e.\ how far the charge carriers drift, thus blurring the optical pump pattern), rather than the wavelength of the mm-wave beam, meaning sub-diffraction limited imaging is possible. Fig.~\ref{fig_mod}(a) shows how the diffusion of charge carriers in the silicon depends on their effective lifetime. Reconstructed images inherit the same level of blurring as the patterning of the mm-wave field has undergone, due to drift of charge carriers. In this case the point spread function of the system is governed by the drift-induced long-range correlations in the mm-wave field patterns used to probe the object \cite{Phillips2016}. The lifetime of the wafer chosen for our experiments is \SI{75}{\micro\second} (float zone silicon bought from Siltronix, dark resistivity = 5\,k$\Omega$, thickness = \SI{390}{\micro\meter}, 120\,nm of SiO$_2$ passivation; see section 2.4 of the supplementary material for more information). A lifetime of \SI{75}{\micro\second} leads to a diffusion length of \SI{350}{\micro\meter}, and a lateral resolution of twice this value i.e.\ \SI{700}{\micro\meter}, which agrees with the resolutions observed in our images. 

There is also scope to achieve lateral resolutions far below our experimental limit by reducing the effective carrier lifetime of the modulator \cite{Hooper2019}. However Fig.~\ref{fig_mod}(a) also shows that, for a given optical pump power, decreasing the carrier lifetime simultaneously decreases the reflection modulation depth, MD = $R_{\textrm{dark}} - R_{\textrm{illum}}$, which depends on the reflected intensity of the mm-wave when the silicon is dark, $R_{\textrm{dark}}$, and illuminated, $R_{\textrm{illum}}$. Ultimately, MD is a function of carrier lifetime, as one can achieve a higher photoconductivity for longer lifetimes (see section 2.5 of the supplementary material). For our experiment, a lifetime of \SI{75}{\micro\second} provided a reasonable trade off between resolution and MD, while higher MD comes at a cost of lower imaging resolution. Conversely, for applications where long imaging times and averaging are possible, one could trade SNR in return for significantly higher imaging resolution.

There are other parameters we can tune to ensure that MD is maximised without compromising on the speed and resolution of imaging - namely the incident angle, wafer thickness and frequency. In Fig.~\ref{fig_mod}(b) we show the numerically calculated MD as a function of incident angle and silicon wafer thickness, on a TPX-silicon-air stack (calculated here for TE polarisation and 140\,GHz with an excitation intensity of 220\,Wm$^{-2}$ and excitation wavelength of 623\,nm for a silicon wafer with effective charge carrier lifetime $\tau_{\textrm{eff}}=\SI{75}{\micro\second}$). The inset shows the reflected mm-wave intensity (for dark and illuminated states), for a \SI{390}{\micro\meter} thick wafer. For a stack of TPX-silicon-air we predict the onset of TIR for a critical angle of $\theta_{\textrm{crit}} = 39.3^{\circ}$. For angles beyond $\theta_{\textrm{crit}}$ the modulation is around an order of magnitude larger than that achieved at normal incidence. The incident angle that provides the largest modulation is near grazing incidence. However, as discussed earlier, angles closer to the critical angle give evanescent fields which penetrate further into low index samples (see Fig.~\ref{fig_schem}(b)), so we opt for an angle of $49^{\circ}$, marked by the red circle, where the penetration length of the field is around \SI{300}{\micro\meter} in air. 

It is also clear from Fig.~\ref{fig_mod}(b) that the modulation is dependent on the thickness of the silicon wafer. This arises due to a cavity resonance in the wafer, as reported in \cite{Hooper2019}, which gives an increased modulation when the wafer thickness is equal to a half-integer multiple of the wavelength in silicon. For a mm-wave frequency of 140 GHz, we choose a wafer that is \SI{390}{\micro\meter} thick; close to the resonance but limited by commercially available silicon wafers.

\section{Conclusions}

In conclusion, we have demonstrated the first computational imaging system in a total internal reflection geometry. This unique combination allows images of sample structure to be reconstructed with sub-wavelength lateral resolution \textit{and} sub-wavelength optical depth sensitivity, which is not possible with other techniques \cite{Wojdyla2013,Johnson2001,Stantchev2020}. In comparison with previously demonstrated THz and mm-wave computational imaging systems \cite{Stantchev2016,Nagatsuma2013,Wojdyla2013}, our design exhibits a high pixels-per-second rate of several kHz, which gives a resolution-dependent frame rate of a few seconds. Our TIR system is also cheaper and more compact than previous work, as it circumvents the need for femtosecond-pulsed lasers.

The penetration depth of the illuminating field is dependent on both incident angle and mm-wave frequency, pointing to an intriguing avenue for future development of the system: by capturing several images at a range of incident angles and frequencies, there is potential to decouple material parameters from layer thickness, and so independently recover these sample properties simultaneously. Section 4.5 of the supplementary material details modelling that demonstrates this principle for some simple cases. In the future it may be possible to extend this approach to recover TIR-based tomographic images of more complex samples that exhibit strong inhomogeneity in all three dimensions. Together, we believe these features point towards many potential applications of mm-wave computational TIR imaging for interrogation of the structure of optically opaque samples in, for example, the medical, art conservation and food industries.

\section*{Funding}
L.E.B and I.R.H. acknowledge financial support from the Engineering and Physical Sciences Research Council of the United Kingdom (EPSRC UK) and QinetiQ Ltd. via the TEAM-A Prosperity Partnership (Grant No. EP/R004781/1). D.B.P acknowledges financial support from the Royal Academy of Engineering (UK) and the European Research Council (Grant No. 804626). E.H. and M.M acknowledge financial support from the EPSRC UK (Grant No. EP/S036466/1). E.H. and D.B.P also acknowledge financial support from the EPSRC UK (Grant No. QuantIC EP/M01326X/1).

\section*{Acknowledgements}
QinetiQ Ltd patent applications relating to a method and apparatus for imaging an object comprising biological material are GB1908140.5 filed 7th June 2019 and GB2003820.4  filed 17th March 2020.

\section*{Disclosures}
L.E.B.: QinetiQ Ltd (F,P), P.K.: QinetiQ Ltd (P), I.R.H.: QinetiQ Ltd (F), S.M.H.: QinetiQ Ltd (P), C.R.L.: QinetiQ Ltd (E,P), E.H.: QinetiQ Ltd (P)


\bibliography{Near-Field}

\end{document}